\documentclass{aastex}

\shorttitle{Optical spectra of southern hemisphere ICRF radio sources}
\shortauthors{Titov et al.}

\begin{document}

\title{Optical Spectra of Candidate Southern Hemisphere International
Celestial Reference Frame (ICRF) radio sources}
\author{O.\ Titov}
\affil{Geoscience Australia, PO Box 378, Canberra, ACT 2601, Australia}
\email{oleg.titov@ga.gov.au}

\author{D.\ L.\ Jauncey}
\affil{CSIRO Astronomy and Space Science, ATNF \& Mount Stromlo Observatory, Cotter Road, Weston, ACT 2611, Australia}

\author{H.\ M.\ Johnston and R.\ W.\ Hunstead}
\affil{Sydney Institute for Astronomy, School of Physics, University of Sydney, NSW 2006, Australia}

\and

\author{L.\ Christensen}
\affil{Technische Universit\"at Munich, Excellence Cluster Universe, 
Boltzmannstr.\ 2, D-85748 Garching}

\begin{abstract} 
 We present the results of spectroscopic observations of the optical
counterparts of 47 southern radio sources from the candidate International
Celestial Reference Catalogue (ICRC), as part of a Very
Long Baseline Interferometry (VLBI)  program to strengthen the celestial
reference frame, especially in the south. The observations were made with
the 3.58-meter European Southern Observatory New Technology Telescope
(NTT). We obtained redshifts for 30 quasars and one radio galaxy, with a
further 7 objects being probable BL Lac objects with featureless spectra.  
Of the remainder, four were clear misidentifications with Galactic stars
and five had low signal-to-noise spectra and could not be classified.  
These results add significantly to the existing data needed to refine the
distribution of source proper motions over the celestial sphere.

\end{abstract}


\section{Introduction}

We are investigating the intrinsic properties and distribution of the
ultra-compact, flat-spectrum radio sources that make up the 
International Very Long Baseline Interferometry (VLBI) Service (IVS)
Reference Catalogue.  Astrometric VLBI measures precise group delays,
difference in arrival times of the radio waves at widely separated
radio telescopes, and thus produces accurate radio positions typically
with milliarcsecond precision. The first systematic astrometric VLBI
program was started in 1979 and, since 2000, IVS has organised a
comprehensive international program of precision position measurements
of about one thousand sources \citep{Schluter2007}.

About 50 radio telescopes have participated in observations over the
30-year IVS history.  Conventional reductions of high-precision
multi-frequency VLBI data are made in the reference system with its origin
at the barycentre of the Solar system \citep{McCarthy2004}.  This
barycentric reference system was adopted by the International Astronomical
Union (IAU) as the International Celestial Reference System (ICRS). By
definition, the axes of the ICRS are fixed by the positions of selected
extragalactic radio sources, and the reference frame representing the
system should show no global rotation with respect to these sources. VLBI
presently represents the most precise, practical and reliable method to
realise such a catalogue.

From 1998 through 2009 the initial International Celestial Reference Frame
(ICRF1) was based on a catalogue of 608 radio VLBI source positions. Of
those, 212 were so-called ``defining'' sources that were used to establish
the orientation of the ICRS axes \citep{Ma1998}.  In 2009 the second
realization of the ICRF (ICRF2) was put forward. ICRF2 contains 295
``defining'' sources of which only 97 were ``defining'' in the ICRF1. The
claimed formal position error of the best observed radio sources in the
ICRF2 is 6--7 $\mu$as, but a more realistic ``inflated'' error is reported
as 41 $\mu$as \citep{Fey2009}.

A separate program is underway to link the optical and radio reference
frames, and one of the important tasks here is to make reliable optical
identifications whenever an optical counterpart can be found. However,
along the Galactic plane the high dust absorption and the high density of
Galactic stars can make identification impossible. Spectroscopic
observations may be able to confirm an identification but more often the
field will be obscured.

The total number of radio sources included in the IVS astrometric program
exceeds 4,000, though only $\sim$ 1,000 sources are observed on a regular
basis. The database of the radio source physical characteristics
\citep{Titov2009} comprises 4261 objects (by August, 2010), mostly
quasars, but there is a serious deficit in the southern hemisphere of
candidate sources for which VLBI observations have been made.  There is
also a significant lack of optical identifications for the existing
candidate radio sources. By July 2010, of the 2211 ICRF2 sources with
measured redshifts, only 781 are in the southern hemisphere and only 129
have declinations south of $-40\degr$. Lack of redshifts in the south can
cause difficulties in the analysis of apparent proper motions of the
reference radio sources \citep{Titov2009}.

Our spectroscopic program therefore focuses on spectroscopic observations
of the IVS radio sources with optical counterparts, especially those in
the south, as well as those with a long observational history.  In this
paper we present spectra of 44 extragalactic objects observed at the
European Southern Observatory's (ESO) 3.58-meter New Technology Telescope
(NTT). We describe the data reduction and analysis procedures in Section 2
and report our results in Section 3.  Throughout the paper we adopt the
source names used by the IVS community; these are similar---and, in 
many cases,
identical---to the original PKS B1950 convention.

\section{Observations}

The observations were carried out in 2010 August at the NTT (Visitor Mode
run 085A-0588 (A)) using the ESO Faint Object Spectrograph and Camera
(EFOSC) system with grism \#13 covering the wavelength range 3685--9315
\AA. The seeing was typically 0.8--1.5 arcseconds but occasionally as high
as 4 arcseconds. The measured spectral resolution was 21 \AA\ FWHM. After
setting up on each target we observed for an initial 15 minutes, followed
by an additional 15 minutes if no obvious emission line was seen on the
first exposure; individual spectra were later combined. Wavelength
calibration was performed using the spectrum of a He/Ne/Ar comparison
lamp, resulting in typical fit errors of 0.5 \AA\ (rms).

Data reduction was performed with the IRAF software suite using standard
procedures for spectral analysis. We removed the bias and pixel-to-pixel
gain variations from each frame and then removed cosmic rays using the
IRAF task {\sc szap}. The separate exposures were then combined and a
single spectrum extracted. We calibrated the resulting one-dimensional
spectrum in wavelength, and flux-calibrated each spectrum by comparing
with the spectrum of a spectrophotometric standard taken with the same
instrumental setup. Because the observing conditions were not photometric,
the flux calibration should be taken as approximate. 


\section{Results}

The outcomes from the spectroscopic observations of 47 targets were as 
follows (Table~1).
\begin{itemize}
\item Emisssion-line redshifts: 31
\item Probable BL Lac objects: 8
\item Galactic stars: 3
\item Spectra with S/N too low to classify: 5
\end{itemize}

Spectra for 44 of the targets are presented in Fig.\ 1, and emission-line 
data for the 31 objects for which redshifts have been determined are given in
Table~2. The radio positions were taken from the ICRF2 catalogue
\citep{Fey2009}. 

Seven objects---\object[IVS
B0122-260]{IVS~B0122$-$260}, B1443$-$162, B1533$-$316, B1633$-$810, 
B2012$-$017, B2053$-$323 and B2254$-$204--- 
were found to have featureless spectra and hence are identified as
probable BL Lac objects.  

Four objects that were found close to the radio position were identified
as stars. Two of these, \object[IVS B1748-253]{IVS~B1748$-$253} and
\object[IVS B1822-173]{IVS~B1822$-$173}, are within $3^\circ$ of the
Galactic plane, are in dense star fields and exhibit extinction in $V$ of
at least 5 magnitudes, so it remains unlikely that their actual optical
counterparts will be detectable.  Radio-optical position differences for
all four objects are given in Table 2, and the identification is rejected
on positional grounds in each case.  More detailed discussion of two
special cases, \object[IVS 1923+210]{IVS~B1923$+$210} and \object[IVS
B2300-307]{IVS~B2300$-$307}, is given in subsections \ref{PKS1923+210} and
\ref{PKS2300-307} respectively.

Finally, there were five targets with low signal/noise spectra that could
not be classified---IVS~B0107$-$610, B1452$-$168, B1647$-$296, B1936$-$623
and B2059$-$786---due either to their faintness or poor observing
conditions.  The field of IVS~B1647$-$296, at $\ell=352^{\circ}.5$,
$b=9^{\circ}.35$, is too crowded to be confident that the object observed
was the correct identification and that the spectrum was not contaminated
by light from a neighbouring star.

\subsection{Notes on individual sources}

\begin{itemize}
 \item \object[IVS B0002-170]{IVS~B0002$-$170} -- reported as a quasar
with redshift $z=0.77539$ (Q=3) in the 6dF survey \citep{Jones2009}; our
redshift of $z=0.7804 \pm 0.0013$ is based on a much higher
signal-to-noise spectrum.

\item \object[IVS B0008-300]{IVS~B0008$-$300} -- strong absorption
feature blueward of the Ly-$\alpha$ emission line. 

\item \object[IVS B0028-396]{IVS~B0028$-$396} -- only two lines were
identified in a noisy spectrum but the redshift is secure. 

\item \object[IVS B0055-059]{IVS~B0055$-$059} -- reported as a quasar
with redshift $z=1.23998$ (Q=2) in the 6dF survey \citep{Jones2009}; our 
redshift of $z=1.2456 \pm 0.0008$ is more reliable because of the higher 
signal-to-noise spectrum. 

\item \object[IVS B1707-038]{IVS~B1707$-$038} -- previously known as a BL
Lac object \citep{Halpern2003}, but weak emission lines were clearly
detected in our spectrum at $z=1.9231$.

\item \object[IVS B1505-304]{IVS~B1505$-$304} -- the highest redshift
object, $z=3.40$, in this sample; strong absorption shortward of 
Ly-$\alpha$, with Lyman limit absorption at or close to the emission 
redshift.

\item \object[IVS B2135-184]{IVS~B2135$-$184} -- original 
galaxy identification by  \citet{Gearhart1972}


\item \object[IVS B2318-087]{IVS~B2318$-$087} -- Faint object, but clear
emission lines of Ly-$\alpha$ and \ion{C}{4} give us confidence that our
redshift of $z=3.1639$ is correct.

\item \object[IVS B2344-514]{IVS~B2344$-$514} -- ICRF2 defining
source. The redshift $z=2.67$ cited by \citet{Fey2009} is not
correct, possibly due to the misidentification of \ion{C}{4} with
Ly-$\alpha$

\end{itemize}

\subsection{IVS~B1923$+$210}\label{PKS1923+210}

The optical counterpart (see Fig 2, object 3) was previously observed at
the Russian 6-meter telescope \citep{Maslennikov2010}, and no emission 
lines were found. For the current campaign we observed all three optical
objects located close to the ICRF2 position at 19h 25m 59.605s +21$^\circ$
06$'$ 26.16$''$ (J2000).  The spectra of objects 1 and 2 both show the
\ion{Ca}{2} triplet at rest, which confirms them as stars.  Object 3,
which has $R \sim 20$ and is coincident with the radio position, shows no
stellar absorption features although the S/N is poor.  There are also no
obvious emission lines.  The combination of a radio-optical coincidence
and featureless spectrum is sufficient to suggest this as a probable BL
Lac object.

\subsection{IVS~B2300$-$307}\label{PKS2300-307}

The object closest to the Tidbinbilla interferometer position was reported
as a star of magnitude 16 by \citet{Jauncey1982} despite close positional
agreement.  Later radio observations confirmed it as a strong source
$\sim$0.5 Jy with a flat spectrum
\citep{Wright1991,Quiniento1993,Reynolds1994}. A VLBI image at 8.4~GHz
published by \citet{Ojha2005} reveals the classical double structure of an
extragalactic radio source, with components separated by $r = 3.3$~mas in
positional angle $124^\circ$. Optical observations reveal a single object
(Fig 3) with a stellar absorption spectrum characteristic of a G star.  
This star is offset by 3.7 arcseconds from the ICRF2 radio position.  The
VLBI radio structure of \object[IVS B2300-307]{IVS~B2300$-$307}
\citep{Ojha2005}, its low radio proper motion, 15 $\mu$as/yr \citep[IVS
data from 2002]{Reynolds1994}, and its high brightness temperature,
$10^{11}$K \citep{Preston1985} all strongly support its extragalactic
nature.  Therefore we conclude that the field of \object[IVS
B2300-307]{IVS~B2300$-$307} is obscured by the foreground star.  Since
there is no sign of an extra image at the precise ICRF2 position, marked
with a cross in Fig.\ 3, we estimate that the optical counterpart probably
has $B\geq 21$.

\section{Comparison of radio and optical positions}

Correct identification of extragalactic radio sources can be also verified
by comparison of their optical and radio coordinates. In our case we
compared radio coordinates from ICRF2 and optical coordinates from the
SuperCOSMOS survey \citep{Hambly2001}. Of the 31 extragalactic sources,
positions for only 25 could be accurately determined;  the remainder were
excluded, either because the images were too faint to be measured reliably
by SuperCOSMOS, or were blended because of low Galactic latitude.  Two
low-redshift objects (IVS~B2135$-$184 and B2211$-$388) were excluded
because their images were clearly non-stellar. We found the mean
difference between the radio and optical position to be $0\farcs{064} \pm
0\farcs{021}$\ with 1-$\sigma$\ rms of 0\farcs{103} in right ascension and
$0\farcs{023} \pm 0\farcs{031}$\ with 1-$\sigma$ rms of 0\farcs{155} in
declination. The scatter in both coordinates is small and supports the
claimed accuracy of SuperCOSMOS \citep{Hambly2001}. While the mean
difference in declination is negligible, the difference in right ascension
marginally exceeds the 3-$\sigma$ level.

A similar radio-optical comparison was carried out for seven of the eight
probable BL Lac objects, excluding IVS~B1923$+$210 which was too faint to
be seen with SuperCOSMOS.  The mean differences were not significant, with
a 1-$\sigma$ scatter of 0\farcs{15} in each coordinate.  This result
supports the identifications and hence their classification as BL Lac
objects.

\section{Summary and conclusion}

We performed spectroscopic observations of the optical counterparts of 47
southern radio sources from the ICRF and measured 31 new redshifts, with
two of them having $z \geq 3$. At least two emission lines were
confidently identified for each target.  Redshifts were determined
principally from isolated lines, with lines that were weak, blended or
affected by strong absorption given lower (or zero) weight.  Eight targets
showed featureless spectra, and were considered as BL Lac objects.

 There were eight ICRF2 defining radio sources in this run. Five of them
---\object[IVS~B1659-621]{IVS~B1659$-$621}, \object[IVS
B1758-651]{B1758$-$651}, \object[IVS~B1815-553]{B1815$-$553},
\object[IVS~B2236-572]{B2236$-$572} and \object[IVS
B2344-514]{B2344$-$514}--- were confirmed as extragalactic sources. Source
\object[IVS~B1631-810]{IVS~B1631$-$810} was confirmed as a BL Lac object
with a featureless spectrum, but two other sources,
\object[IVS~B0107-610]{IVS~B0107$-$610} and \object[IVS
B1443-162]{IVS~B1443$-$162}, were observed in poor weather conditions, and
no emission lines could be confidently identified in their spectra.

\acknowledgments

This paper is based on observations collected at the European
Organisation for Astronomical Research in the Southern Hemisphere,
Chile, programme 058.A-0855(A). Two of us, Titov and Jauncey, were
supported by a travel grant from the Australian Nuclear Science 
Technology Organisation (ANSTO) in their Access to Major
Research Facilities Program (AMRFP) (reference number AMRFP
10/11-O-04) to travel to La Silla. The paper is published with the
permission of the CEO, Geoscience Australia.

{\it Facilities:} \facility{NTT (GMOS)}

\clearpage

\begin{deluxetable}{llcccccc}
\tablecaption{Outcomes of the spectroscopic observations}   
\tablewidth{0pt}
\tablehead{
  \colhead{Source} &
\colhead{RA (J2000)\tablenotemark{a}} & \colhead{Dec 
(J2000)\tablenotemark{a}} &
  \colhead{Redshift}
}
\startdata
IVS B0002$-$170 & 00 05 17.933 & $-$16 48 04.678 & 0.7804 $\pm$ 0.0013  \\
IVS B0008$-$300 & 00 10 45.177 & $-$29 45 13.177 & 2.3439 $\pm$ 0.0052  \\
IVS B0028$-$396 & 00 31 24.331 & $-$39 22 49.391 & 1.2978 $\pm$ 0.0023  \\
IVS B0034$-$220 & 00 37 14.825 & $-$21 45 24.714 & 2.5133 $\pm$ 0.0021  \\
IVS B0055$-$059 & 00 58 05.066 & $-$05 39 52.277 & 1.2456 $\pm$ 0.0008  \\
IVS B0107$-$610 & 01 09 15.475 & $-$60 49 48.460 & low S/N \\
IVS B0110$-$668 & 01 12 18.912 & $-$66 34 45.187 & 1.1888 $\pm$ 0.0021  \\
IVS B0122$-$260 & 01 25 18.837 & $-$25 49 04.390 & BL Lac \\
IVS B0221$-$171 & 02 23 43.763 & $-$16 56 37.701 & 1.0152 $\pm$ 0.0017  \\
IVS B1443$-$162 & 14 45 53.376 & $-$16 29 01.619 & BL Lac \\
IVS B1452$-$168 & 14 55 02.811 & $-$17 00 13.953 & low S/N \\
IVS B1505$-$304 & 15 08 52.993 & $-$30 36 29.430 & 3.3684 $\pm$ 0.0023  \\   
IVS B1511$-$476 & 15 14 40.024 & $-$47 48 29.858 & 1.5512 $\pm$ 0.0018  \\
IVS B1533$-$316 & 15 36 54.498 & $-$31 51 15.135 & BL Lac \\
IVS B1633$-$810 & 16 42 57.346 & $-$81 08 35.070 & BL Lac \\
IVS B1635$-$141 & 16 38 45.284 & $-$14 15 50.237 & 0.2575 $\pm$ 0.0004  \\
IVS B1647$-$296 & 16 50 39.544 & $-$29 43 46.955 & low S/N \\
IVS B1659$-$621 & 17 03 36.541 & $-$62 12 40.008 & 1.7547 $\pm$ 0.0012  \\
IVS B1707$-$038 & 17 10 17.205 & $-$03 55 50.128 & 1.9231 $\pm$ 0.0017  \\
IVS B1726$-$038 & 17 28 50.235 & $-$03 50 50.436 & 0.6617 $\pm$ 0.0003  \\
IVS B1748$-$253 & 17 51 51.263 & $-$25 24 00.064 & star \\
IVS B1758$-$651 & 18 03 23.496 & $-$65 07 36.761 & 1.1991 $\pm$ 0.0006  \\
IVS B1815$-$553 & 18 19 45.399 & $-$55 21 20.745 & 1.6292 $\pm$ 0.0013  \\
IVS B1822$-$173 & 18 25 36.532 & $-$17 18 49.848 & star \\
IVS B1852$-$534 & 18 57 00.452 & $-$53 25 00.356 & 0.7779 $\pm$ 0.0007  \\
IVS B1923+210   & 19 25 59.605 &   +21 06 26.162 & BL Lac \\
IVS B1928$-$698 & 19 33 31.159 & $-$69 42 58.914 & 1.4807 $\pm$ 0.0020  \\
IVS B1936$-$623 & 19 41 21.769 & $-$62 11 21.056 & low S/N \\
IVS B2012$-$017 & 20 15 15.158 & $-$01 37 32.560 & BL Lac \\
IVS B2053$-$323 & 20 56 25.070 & $-$32 08 47.801 & BL Lac \\
IVS B2059$-$786 & 21 05 44.961 & $-$78 25 34.547 & low S/N \\
IVS B2107$-$105 & 21 10 00.978 & $-$10 20 57.319 & 2.5004 $\pm$ 0.0012  \\
IVS B2117$-$614 & 21 21 04.074 & $-$61 11 24.624 & 1.0168 $\pm$ 0.0011  \\
IVS B2135$-$184 & 21 38 41.928 & $-$18 10 44.371 & 0.1887 $\pm$ 0.0001  \\
IVS B2158$-$167 & 22 00 54.878 & $-$16 32 32.701 & 0.8355 $\pm$ 0.0024  \\
IVS B2211$-$388 & 22 14 38.569 & $-$38 35 45.008 & 0.3888 $\pm$ 0.0004  \\
IVS B2220$-$163 & 22 23 41.172 & $-$16 07 05.188 & 0.8811 $\pm$ 0.0002  \\
IVS B2234$-$253 & 22 37 18.355 & $-$25 06 32.519 & 1.2788 $\pm$ 0.0011  \\
IVS B2236$-$572 & 22 39 12.075 & $-$57 01 00.839 & 0.5686 $\pm$ 0.0023  \\
IVS B2239$-$631 & 22 43 07.839 & $-$62 50 57.322 & 0.3924 $\pm$ 0.0005  \\
IVS B2254$-$204 & 22 56 41.208 & $-$20 11 40.510 & BL Lac \\
IVS B2300$-$307 & 23 03 05.821 & $-$30 30 11.473 & star \\
IVS B2318$-$087 & 23 21 18.250 & $-$08 27 21.521 & 3.1639 $\pm$ 0.0033  \\
IVS B2321$-$065 & 23 23 39.113 & $-$06 17 59.238 & 2.1440 $\pm$ 0.0020  \\
IVS B2327$-$459 & 23 30 37.680 & $-$45 39 58.101 & 0.4471 $\pm$ 0.0004  \\
IVS B2344$-$514 & 23 47 19.864 & $-$51 10 36.065 & 1.7502 $\pm$ 0.0060  \\
IVS B2354$-$251 & 23 57 23.850 & $-$24 51 03.163 & 1.6137 $\pm$ 0.0014  \\
\enddata
\tablenotetext{a}{ICRF2 radio position}
\end{deluxetable}

\begin{deluxetable}{llcccccc}
\tabletypesize{\small}
\tablecaption{Observed emission lines}
\label{tab:linelist}
\tablewidth{0pt}
\tablehead{
  \colhead{Source} & 
  \colhead{Line} & \colhead{$\lambda_\mathrm{rest}$\ (\AA)} &
  \colhead{$\lambda_\mathrm{obs}$\ (\AA)\tablenotemark{a}} &
  \colhead{$z$} & \colhead{mean $z$}
}
\startdata
IVS B0002$-$170 & [\ion{O}{2}]   & 2470.2 & 4398.0\phm{:} & 0.7804 & 0.7804 $\pm$ 0.0013 \\
                & \ion{Mg}{2}    & 2797.9 & 4986.9\phm{:} & 0.7824 & \\
                & [\ion{Ne}{5}]  & 3425.5 & 6086.4\phm{:} & 0.7768 & \\
                & H$\gamma$      & 4340.5 & 7715.7:       & 0.7776 & \\
                & [\ion{O}{3}]   & 4363.2 & 7765.4:       & 0.7797 & \\
                & H$\beta$       & 4861.3 & 8673.7\phm{:} & 0.7842 & \\
                & [\ion{O}{3}]   & 4958.9 & 8840.4:       & 0.7827 & \\
                & [\ion{O}{3}]   & 5006.8 & 8904.4\phm{:} & 0.7784 & \\

IVS B0008$-$300 & Ly$\alpha$     & 1215.7 & 4090.0:       & 2.3644 & 2.3439 $\pm$ 0.0052 \\
                & \ion{N}{5}     & 1240.1 & 4176.4:       & 2.3677 & \\
                & \ion{Si}{4}    & 1396.8 & 4683.9\phm{:} & 2.3534 & \\
                & \ion{C}{4}     & 1549.1 & 5167.0\phm{:} & 2.3356 & \\
                & \ion{C}{3}]    & 1908.7 & 6380.1\phm{:} & 2.3426 & \\

IVS B0028$-$396 & \ion{C}{3}]    & 1908.7 & 4381.4\phm{:} & 1.2955 & 1.2978 $\pm$ 0.0023 \\
                & \ion{Mg}{2} \phm{:}   & 2797.9 & 6435.4 & 1.3001 & \\

IVS B0034$-$220 & Ly$\alpha$     & 1215.7 & 4276.2\phm{:} & 2.5176 & 2.5133 $\pm$ 0.0021 \\
                & \ion{N}{5}     & 1240.1 & 4355.3:       & 2.5120 & \\
                & \ion{Si}{4}    & 1396.8 & 4909.6\phm{:} & 2.5150 & \\
                & \ion{C}{4}     & 1549.1 & 5441.7\phm{:} & 2.5129 & \\
                & \ion{C}{3}]    & 1908.7 & 6695.6\phm{:} & 2.5079 & \\

IVS B0055$-$059 & \ion{C}{3}]    & 1908.7 & 4287.5\phm{:} & 1.2463 & 1.2456 $\pm$ 0.0008 \\
                & \ion{Mg}{2}    & 2797.9 & 6284.2\phm{:} & 1.2460 & \\
                & [\ion{O}{2}]   & 3726.8 & 8364.9\phm{:} & 1.2446 & \\

IVS B0110$-$668 & \ion{C}{3}]    & 1908.7 & 4174.2\phm{:} & 1.1869 & 1.1888 $\pm$ 0.0021 \\
                & \ion{Mg}{2}    & 2797.9 & 6129.1\phm{:} & 1.1906 & \\

IVS B0221$-$171 & \ion{C}{3}]    & 1908.7 & 3840.8\phm{:} & 1.0122 & 1.0152 $\pm$ 0.0017 \\
                & \ion{Mg}{2}    & 2797.9 & 5639.3\phm{:} & 1.0155 & \\
                & H$\gamma$      & 4340.5 & 8758.9\phm{:} & 1.0180 & \\

IVS B1505$-$304 & Ly lim         & 911.5  & 4031.2:       & 3.4226 & 3.3684 $\pm$ 0.0023 \\
                & Ly$\beta$      & 1025.7 & 4524.4:       & 3.4110 & \\
                & Ly$\alpha$     & 1215.7 & 5412.6:       & 3.4524 & \\
                & \ion{N}{5}     & 1240.1 & 5466.2:       & 3.4077 & \\
                & \ion{Si}{4}    & 1396.8 & 6095.4\phm{:} & 3.3639 & \\
                & \ion{C}{4}     & 1549.1 & 6770.2\phm{:} & 3.3705 & \\
                & \ion{C}{3}]    & 1908.7 & 8342.4\phm{:} & 3.3707 & \\

IVS B1511$-$476 & \ion{C}{4}     & 1549.1 & 3957.3\phm{:} & 1.5546 & 1.5512 $\pm$ 0.0018 \\
                & \ion{He}{2}    & 1640.4 & 4185.6\phm{:} & 1.5515 & \\
                & \ion{C}{3}]    & 1908.7 & 4859.9\phm{:} & 1.5461 & \\
                & \ion{Mg}{2}    & 2797.9 & 7142.2\phm{:} & 1.5527 & \\

IVS B1635$-$141 & [\ion{Ne}{5}]  & 3345.4 & 4206.9\phm{:} & 0.2575 & 0.2575 $\pm$ 0.0004 \\
                & [\ion{Ne}{5}]  & 3425.5 & 4315.8\phm{:} & 0.2599 & \\
                & [\ion{O}{2}]   & 3726.0 & 4682.7\phm{:} & 0.2568 & \\
                & [\ion{Ne}{3}]  & 3869.1 & 4868.1\phm{:} & 0.2582 & \\
                & [\ion{Ne}{3}]  & 3967.8 & 4987.0\phm{:} & 0.2569 & \\
                & H$\delta$      & 4101.7 & 5155.7\phm{:} & 0.2570 & \\
                & H$\gamma$      & 4340.5 & 5466.8\phm{:} & 0.2595 & \\
                & H$\beta$       & 4861.3 & 6111.1\phm{:} & 0.2571 & \\
                & [\ion{O}{3}]   & 4958.9 & 6229.7\phm{:} & 0.2563 & \\
                & [\ion{O}{3}]   & 5006.8 & 6289.7\phm{:} & 0.2562 & \\
                & H$\alpha$      & 6562.8 & 8233.8:       & 0.2546 & \\
                & [\ion{N}{2}]   & 6583.5 & 8267.3:       & 0.2558 & \\

IVS B1659$-$621 & \ion{Si}{4}    & 1396.8 & 3845.5\phm{:} & 1.7532 & 1.7547 $\pm$ 0.0012 \\
                & \ion{C}{4}     & 1549.1 & 4269.0\phm{:} & 1.7559 & \\
                & \ion{C}{3}]    & 1908.7 & 5253.0\phm{:} & 1.7521 & \\
                & \ion{Mg}{2}    & 2797.9 & 7716.0\phm{:} & 1.7578 & \\

IVS B1707$-$038 & \ion{C}{4}     & 1549.1 & 4523.0\phm{:} & 1.9199 & 1.9231 $\pm$ 0.0017 \\
                & \ion{C}{3}]    & 1908.7 & 5584.3\phm{:} & 1.9256 & \\
                & \ion{Mg}{2}    & 2797.9 & 8180.6\phm{:} & 1.9238 & \\

IVS B1726$-$038 & \ion{Mg}{2}    & 2797.9 & 4647.3\phm{:} & 0.6610 & 0.6617 $\pm$ 0.0003 \\
                & [\ion{Ne}{5}]  & 3345.4 & 5559.8\phm{:} & 0.6619 & \\
                & [\ion{Ne}{5}]  & 3425.5 & 5689.8\phm{:} & 0.6610 & \\
                & [\ion{O}{2}]   & 3726.8 & 6195.7\phm{:} & 0.6625 & \\
                & [\ion{Ne}{3}]  & 3869.1 & 6428.5\phm{:} & 0.6615 & \\
                & [\ion{Ne}{3}]  & 3967.8 & 6590.6\phm{:} & 0.6610 & \\
                & H$\gamma$      & 4340.5 & 7212.0\phm{:} & 0.6616 & \\
                & H$\beta$       & 4861.3 & 8077.8\phm{:} & 0.6616 & \\
                & [\ion{O}{3}]   & 4958.9 & 8246.2\phm{:} & 0.6629 & \\
                & [\ion{O}{3}]   & 5006.8 & 8321.9\phm{:} & 0.6621 & \\

IVS B1758$-$651 & \ion{C}{3}]    & 1908.7 & 4199.1\phm{:} & 1.2000 & 1.1991 $\pm$ 0.0006 \\
                & \ion{C}{2}]    & 2326.9 & 5117.0\phm{:} & 1.1990 & \\
                & [Ne IV]        & 2423.8 & 5327.0\phm{:} & 1.1977 & \\
                & \ion{Mg}{2}    & 2797.9 & 6156.8\phm{:} & 1.2005 & \\
                & [\ion{Ne}{5}]  & 3345.4 & 7331.9:       & 1.1916 & \\
                & [\ion{Ne}{5}]  & 3425.5 & 7527.8\phm{:} & 1.1976 & \\
                & [\ion{O}{2}]   & 3726.8 & 8197.9\phm{:} & 1.1997 & \\

IVS B1815$-$553 & \ion{C}{4}     & 1549.1 & 4070.3\phm{:} & 1.6276 & 1.6292 $\pm$ 0.0013 \\
                & \ion{C}{3}]    & 1908.7 & 5016.8\phm{:} & 1.6284 & \\
                & \ion{C}{2}]    & 2326.9 & 6123.7\phm{:} & 1.6317 & \\
                & \ion{Mg}{2}    & 2797.9 & 7384.2:       & 1.6391 & \\

IVS B1852$-$534 & \ion{Mg}{2}    & 2797.9 & 4973.4\phm{:} & 0.7775 & 0.7779 $\pm$ 0.0007 \\
                & [\ion{O}{2}]   & 3726.8 & 6628.4\phm{:} & 0.7786 & \\
                & H$\delta$      & 4101.7 & 7280.6\phm{:} & 0.7750 & \\
                & H$\gamma$      & 4340.5 & 7715.9\phm{:} & 0.7777 & \\
                & [\ion{O}{3}]   & 4958.9 & 8830.3\phm{:} & 0.7807 & \\
                & [\ion{O}{3}]   & 5006.8 & 8900.6\phm{:} & 0.7777 & \\

IVS B1928$-$698 & \ion{C}{4}     & 1549.1 & 3836.7\phm{:} & 1.4768 & 1.4807 $\pm$ 0.0020 \\
                & \ion{C}{3}]    & 1908.7 & 4730.3\phm{:} & 1.4782 & \\
                & \ion{C}{2}]    & 2326.9 & 5755.2:       & 1.4733 & \\
                & \ion{Mg}{2}    & 2797.9 & 6952.9\phm{:} & 1.4850 & \\
                & [\ion{Ne}{5}]  & 3345.4 & 8306.2\phm{:} & 1.4829 & \\

IVS B2107$-$105 & Ly$\alpha$     & 1215.7 & 4258.3\phm{:} & 2.5028 & 2.5004 $\pm$ 0.0012 \\
                & \ion{N}{5}     & 1240.1 & 4347.6:       & 2.5057 & \\
                & \ion{Si}{4}    & 1396.8 & 4889.9\phm{:} & 2.5009 & \\
                & \ion{C}{4}     & 1549.1 & 5418.5\phm{:} & 2.4979 & \\
                & \ion{C}{3}]    & 1908.7 & 6680.4\phm{:} & 2.4999 & \\

IVS B2117$-$614 & \ion{C}{3}]    & 1908.7 & 3848.3\phm{:} & 1.0161 & 1.0168 $\pm$ 0.0011 \\
                & \ion{C}{2}]    & 2326.9 & 4685.5\phm{:} & 1.0136 & \\
                & \ion{Mg}{2}    & 2797.9 & 5651.2\phm{:} & 1.0198 & \\
                & [\ion{O}{2}]   & 3726.8 & 7514.2\phm{:} & 1.0163 & \\
                & H$\gamma$      & 4340.5 & 8760.6\phm{:} & 1.0184 & \\

IVS B2135$-$184 & [\ion{O}{2}]   & 3727.7 & 4430.9\phm{:} & 0.1887 & 0.1887 $\pm$ 0.0001 \\
                & [\ion{Ne}{3}]  & 3869.1 & 4598.4\phm{:} & 0.1885 & \\
                & Ca H           & 3934.8 & 4680.8\phm{:} & 0.1896 & \\
                & Ca K           & 3969.6 & 4718.1\phm{:} & 0.1886 & \\
                & H$\gamma$      & 4340.5 & 5158.2\phm{:} & 0.1884 & \\
                & H$\beta$       & 4861.3 & 5778.3\phm{:} & 0.1886 & \\
                & [\ion{O}{3}]   & 4958.9 & 5889.4\phm{:} & 0.1876 & \\
                & [\ion{O}{3}]   & 5006.8 & 5949.7\phm{:} & 0.1883 & \\
                & Na D           & 5892.9 & 7005.7\phm{:} & 0.1888 & \\
                & H$\alpha$      & 6562.8 & 7805.7\phm{:} & 0.1894 & \\
                & [\ion{S}{2}]   & 6716.4 & 7986.3\phm{:} & 0.1891 & \\

IVS B2158$-$167 & \ion{Mg}{2}    & 2797.9 & 5142.3\phm{:} & 0.8379 & 0.8355 $\pm$ 0.0024 \\
                & [\ion{O}{3}]   & 5006.8 & 9178.2\phm{:} & 0.8331 & \\

IVS B2211$-$388 & [\ion{O}{2}]   & 3727.7 & 5180.5\phm{:} & 0.3897 & 0.3888 $\pm$ 0.0004 \\
                & H$\beta$       & 4861.3 & 6747.3:       & 0.3880 & \\
                & [\ion{O}{3}]   & 4958.9 & 6886.8:       & 0.3888 & \\
                & [\ion{O}{3}]   & 5006.8 & 6954.6\phm{:} & 0.3890 & \\
                & H$\alpha$      & 6562.8 & 9108.5\phm{:} & 0.3879 & \\
                & [\ion{N}{2}]   & 6583.5 & 9140.3\phm{:} & 0.3884 & \\

IVS B2220$-$163 & \ion{C}{2}]    & 2326.9 & 4385.4:       & 0.8846 & 0.8811 $\pm$ 0.0002 \\
                & \ion{Mg}{2}    & 2797.9 & 5262.7\phm{:} & 0.8809 & \\
                & [\ion{O}{2}]   & 3726.8 & 7011.9\phm{:} & 0.8815 & \\
                & [\ion{Ne}{3}]  & 3869.1 & 7277.4\phm{:} & 0.8809 & \\

IVS B2234$-$253 & \ion{C}{3}]    & 1908.7 & 4347.3\phm{:} & 1.2776 & 1.2788 $\pm$ 0.0011 \\
                & \ion{Mg}{2}    & 2797.9 & 6379.3\phm{:} & 1.2800 & \\

IVS B2236$-$572 & \ion{Mg}{2}    & 2797.9 & 4381.2\phm{:} & 0.5659 & 0.5686 $\pm$ 0.0023 \\
                & [\ion{Ne}{5}]  & 3425.5 & 5388.8\phm{:} & 0.5731 & \\
                & [\ion{O}{2}]   & 3726.8 & 5838.7\phm{:} & 0.5667 & \\

IVS B2239$-$631 & \ion{Mg}{2}    & 2797.9 & 3901.5\phm{:} & 0.3944 & 0.3924 $\pm$ 0.0005 \\
                & [\ion{Ne}{5}]  & 3345.4 & 4649.6:       & 0.3898 & \\
                & [\ion{Ne}{5}]  & 3425.5 & 4774.9\phm{:} & 0.3939 & \\
                & [\ion{Ne}{3}]  & 3869.1 & 5385.5\phm{:} & 0.3919 & \\
                & H$\gamma$      & 4340.5 & 6056.1:       & 0.3953 & \\
                & H$\beta$       & 4861.3 & 6767.9\phm{:} & 0.3922 & \\
                & [\ion{O}{3}]   & 4958.9 & 6899.8\phm{:} & 0.3914 & \\
                & [\ion{O}{3}]   & 5006.8 & 6968.1\phm{:} & 0.3917 & \\
                & H$\alpha$      & 6562.8 & 9129.7\phm{:} & 0.3911 & \\

IVS B2318$-$087 & Ly$\alpha$     & 1215.7 & 5065.5\phm{:} & 3.1668 & 3.1639 $\pm$ 0.0033 \\
                & \ion{C}{4}     & 1549.1 & 6445.6\phm{:} & 3.1610 & \\
                & \ion{C}{3}]    & 1908.7 & 7976.2:       & 3.1788 & \\

IVS B2321$-$065 & Ly$\alpha$     & 1215.7 & 3820.9\phm{:} & 2.1430 & 2.1440 $\pm$ 0.0020 \\
                & \ion{N}{5}     & 1240.1 & 3901.7:       & 2.1462 & \\
                & \ion{Si}{4}    & 1396.8 & 4414.6:       & 2.1606 & \\
                & \ion{C}{4}     & 1549.1 & 4865.6\phm{:} & 2.1410 & \\
                & \ion{C}{3}]    & 1908.7 & 5970.9:       & 2.1282 & \\
                & \ion{Mg}{2}    & 2797.9 & 8807.9\phm{:} & 2.1480 & \\

IVS B2327$-$459 & \ion{Mg}{2}    & 2797.9 & 4051.8\phm{:} & 0.4482 & 0.4471 $\pm$ 0.0004 \\
                & [\ion{O}{2}]   & 3726.8 & 5392.7\phm{:} & 0.4470 & \\
                & H$\beta$       & 4861.3 & 7033.7\phm{:} & 0.4469 & \\
                & [\ion{O}{3}]   & 4958.9 & 7169.0:       & 0.4457 & \\
                & [\ion{O}{3}]   & 5006.8 & 7242.4\phm{:} & 0.4465 & \\

IVS B2344$-$514 & \ion{Si}{4}    & 1396.8 & 3873.2:       & 1.7730 & 1.7502 $\pm$ 0.0060 \\
                & \ion{C}{4}     & 1549.1 & 4250.8\phm{:} & 1.7441 & \\
                & \ion{C}{3}]    & 1908.7 & 5261.0\phm{:} & 1.7563 & \\
                & \ion{Mg}{2}    & 2797.9 & 7700.3:       & 1.7521 & \\

IVS B2354$-$251 & \ion{C}{4}     & 1549.1 & 4045.2\phm{:} & 1.6114 & 1.6137 $\pm$ 0.0014 \\
                & \ion{C}{3}]    & 1908.7 & 4992.4\phm{:} & 1.6156 & \\
                & \ion{Mg}{2}    & 2797.9 & 7314.0\phm{:} & 1.6141 & \\
\enddata
\tablenotetext{a}{A colon after the observed wavelength indicates that
the line position was uncertain, because of blending, proximity to a
noise spike, or associated absorption, and hence was not used to
determine the redshift}
\end{deluxetable}

\clearpage

\begin{table}
\begin{center}
\caption{Comparison between radio and optical positions for the stellar 
misidentifications}
\begin{tabular}{lcccccccc}
\tableline\tableline
  Source & \multicolumn{2}{c}{ICRF2 position} &
           \multicolumn{2}{c}{Optical position} & 
  $\Delta \alpha$ & $\Delta \delta$ & $\sigma_\mathrm{opt}$ \\
    & \multicolumn{2}{c}{J2000} & \multicolumn{2}{c}{J2000} & 
      \multicolumn{2}{c}{ICRF2$-$optical} & \\
    & $^\mathrm{h}$~~$^\mathrm{m}$~~$^\mathrm{s}$ &
    $\arcdeg$~~$\arcmin$~~$\arcsec$ & $^\mathrm{s}$ & $\arcsec$ &
    $\arcsec$ & $\arcsec$ & $\arcsec$ \\
\tableline
IVS B1748$-$253 & 17 51 51.263 & $-$25 24 00.06 & 51.104 & 01.44 &    2.15 &   1.38 &  0.29 \\
IVS B1822$-$173 & 18 25 36.532 & $-$17 18 49.85 & 36.566 & 50.57 & $-$0.49 &   0.72 &  0.21 \\
IVS B1923+210   & 19 25 59.605 & $+$21 06 26.16 & 59.662 & 24.37 & $-$0.80 &$-$1.79 &  0.22 \\
IVS B2300$-$307 & 23 03 05.821 & $-$30 30 11.47 & 05.897 & 15.06 & $-$0.98 &   3.59 &  0.15 \\
\tableline
\end{tabular}
\end{center}
\end{table}

\begin{figure}
\epsscale{1.0}
\plottwo{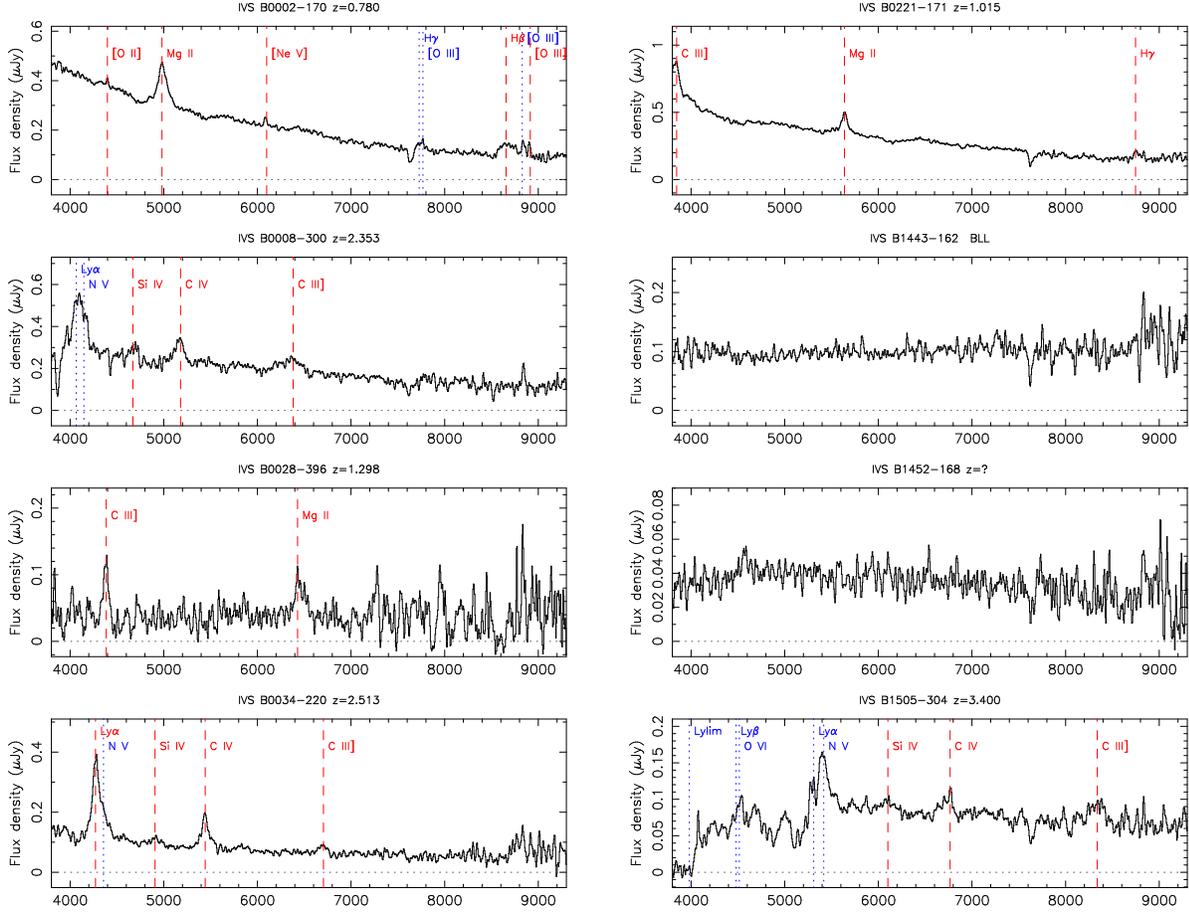}{allspec3.ps}
\caption{Optical spectra for 44 IVS targets, including 31
  emission-line objects listed in Table~2, eight possible BL Lac
  objects, and five low signal-to-noise spectra that could not be
  classified.  Dashed lines indicate emission lines used for redshift
  determination; dotted lines indicate features that were not used.}
\end{figure}

\begin{figure}
\figurenum{1 (continued)}
\plottwo{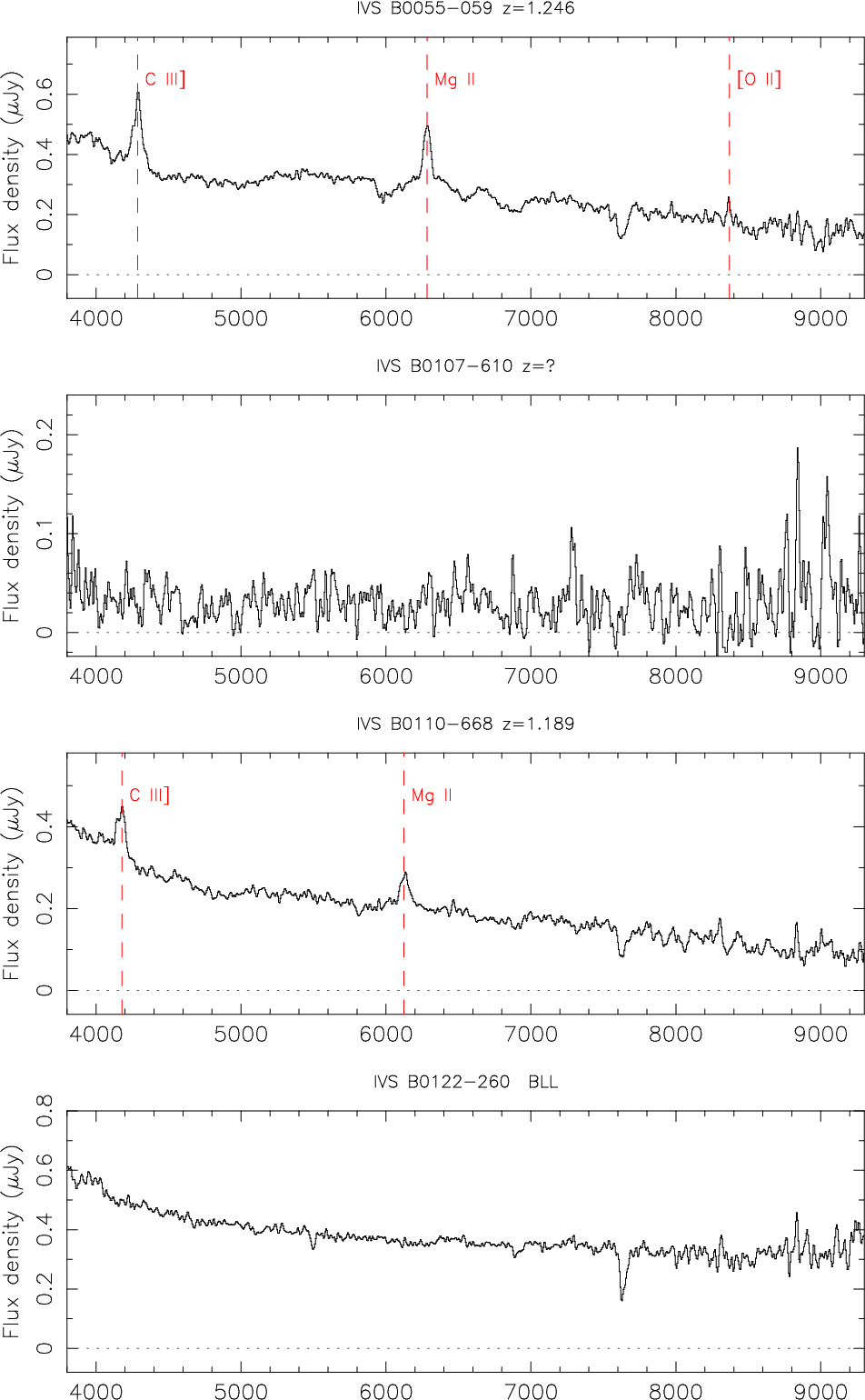}{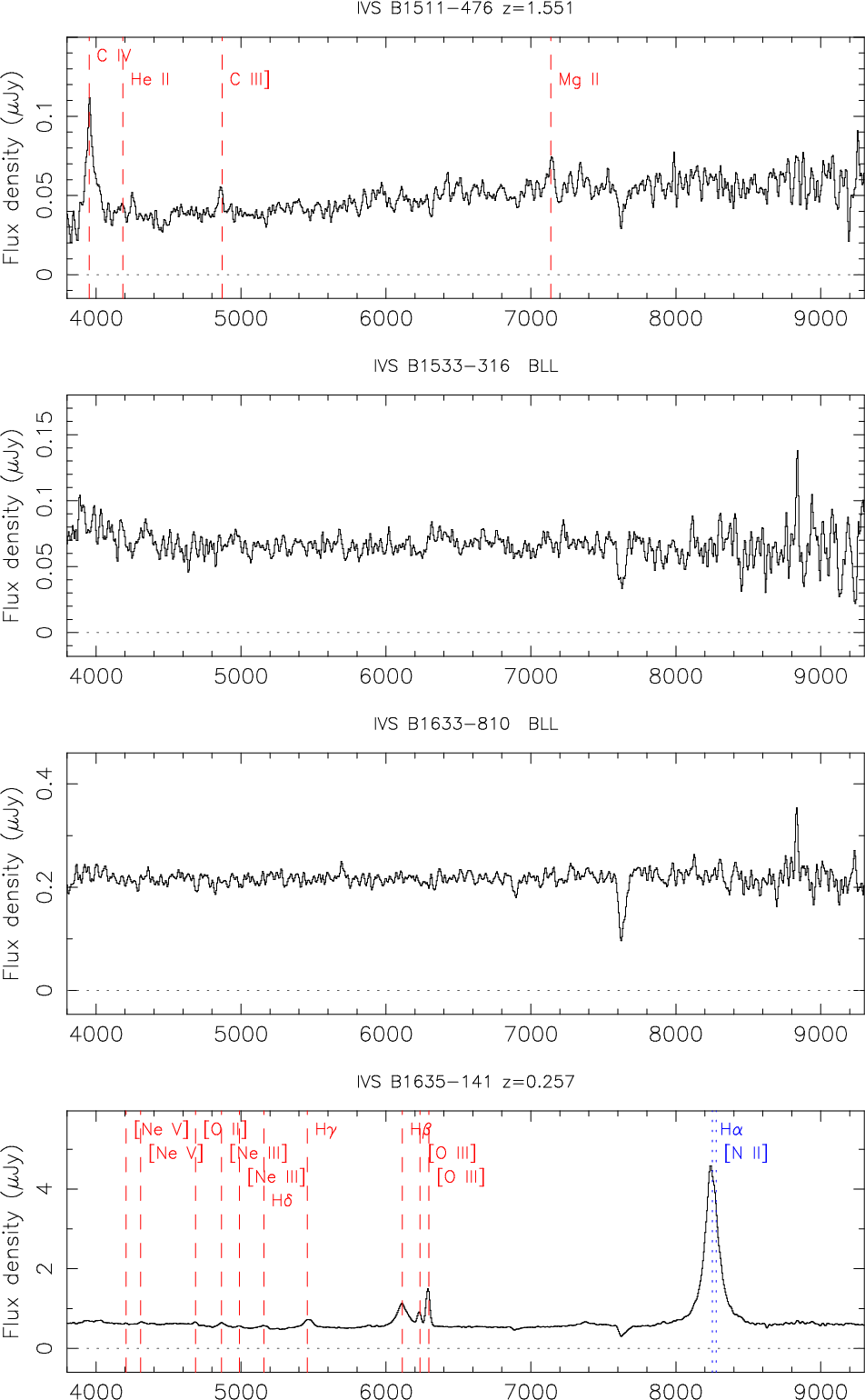}
\caption{}
\end{figure}

\begin{figure}
\figurenum{1 (continued)}
\plottwo{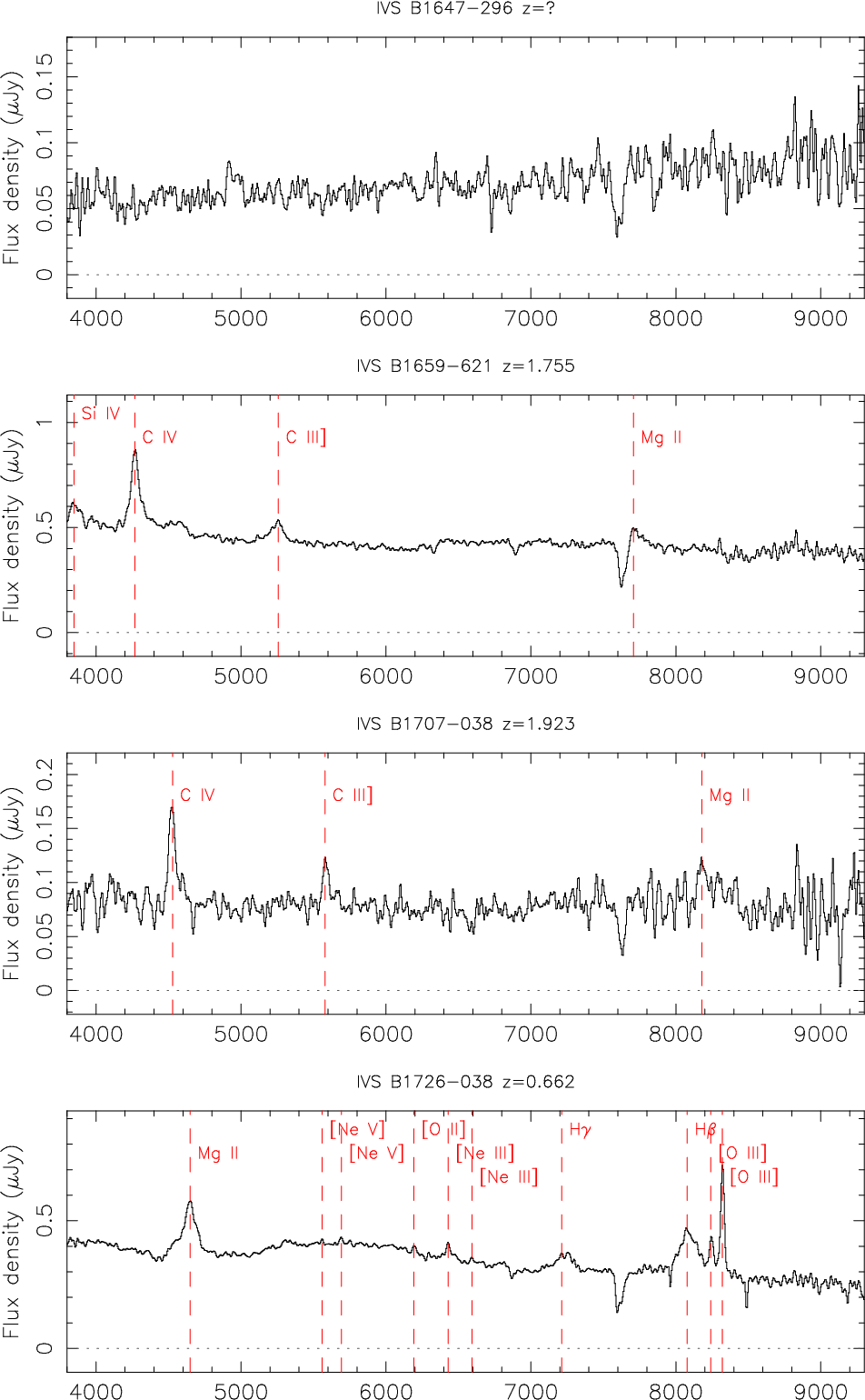}{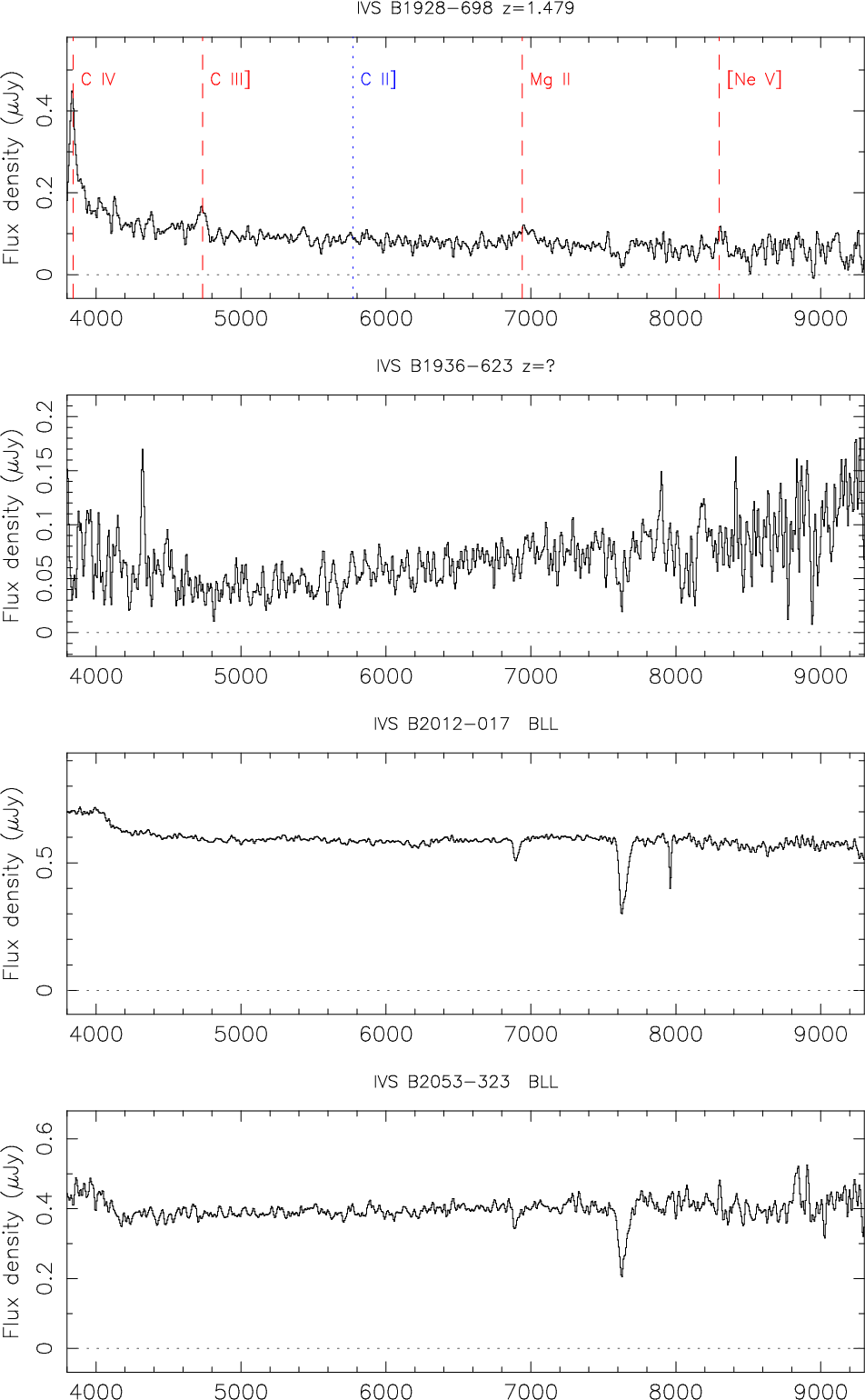}
\caption{}
\end{figure}

\begin{figure}
\figurenum{1 (continued)}
\plottwo{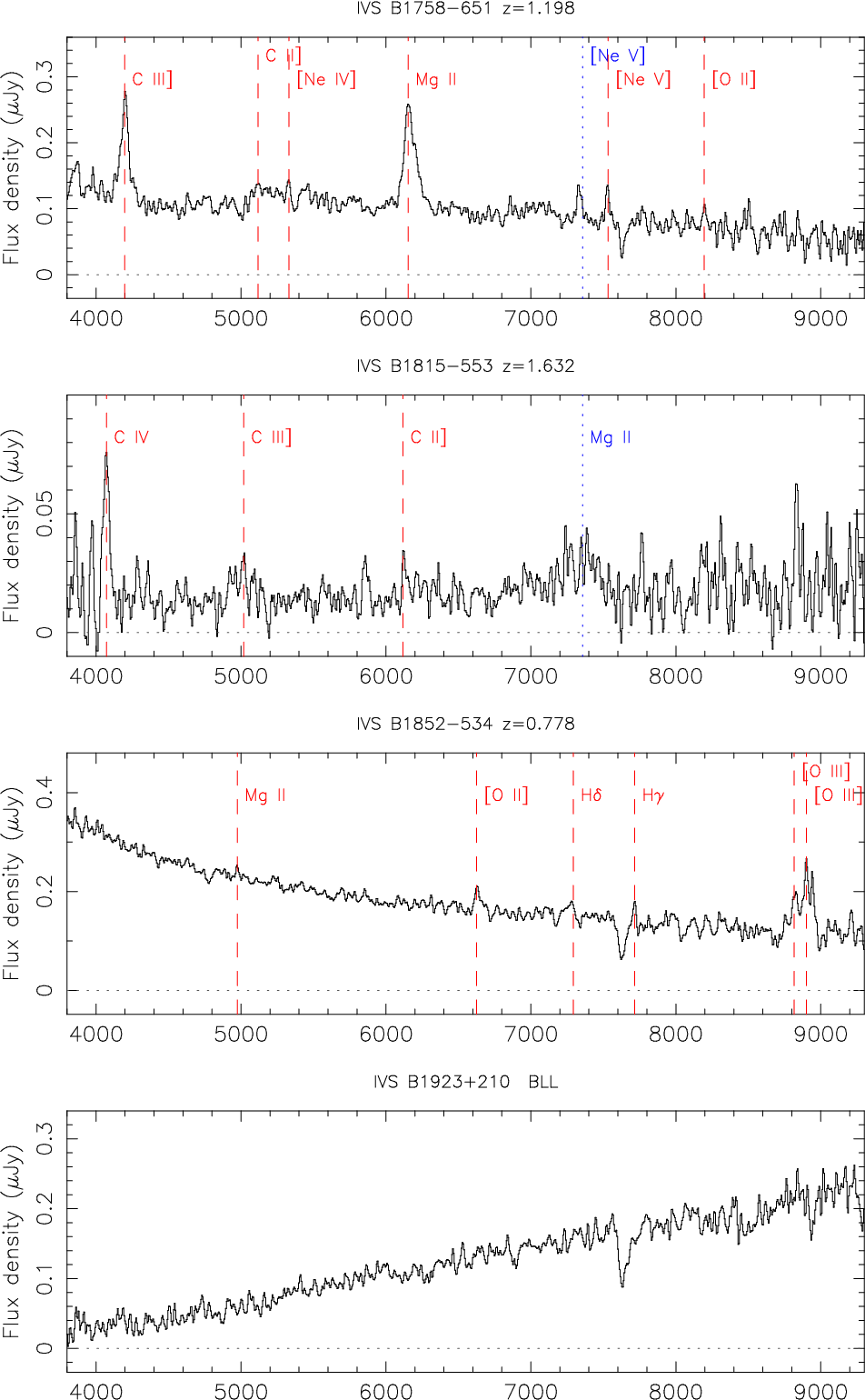}{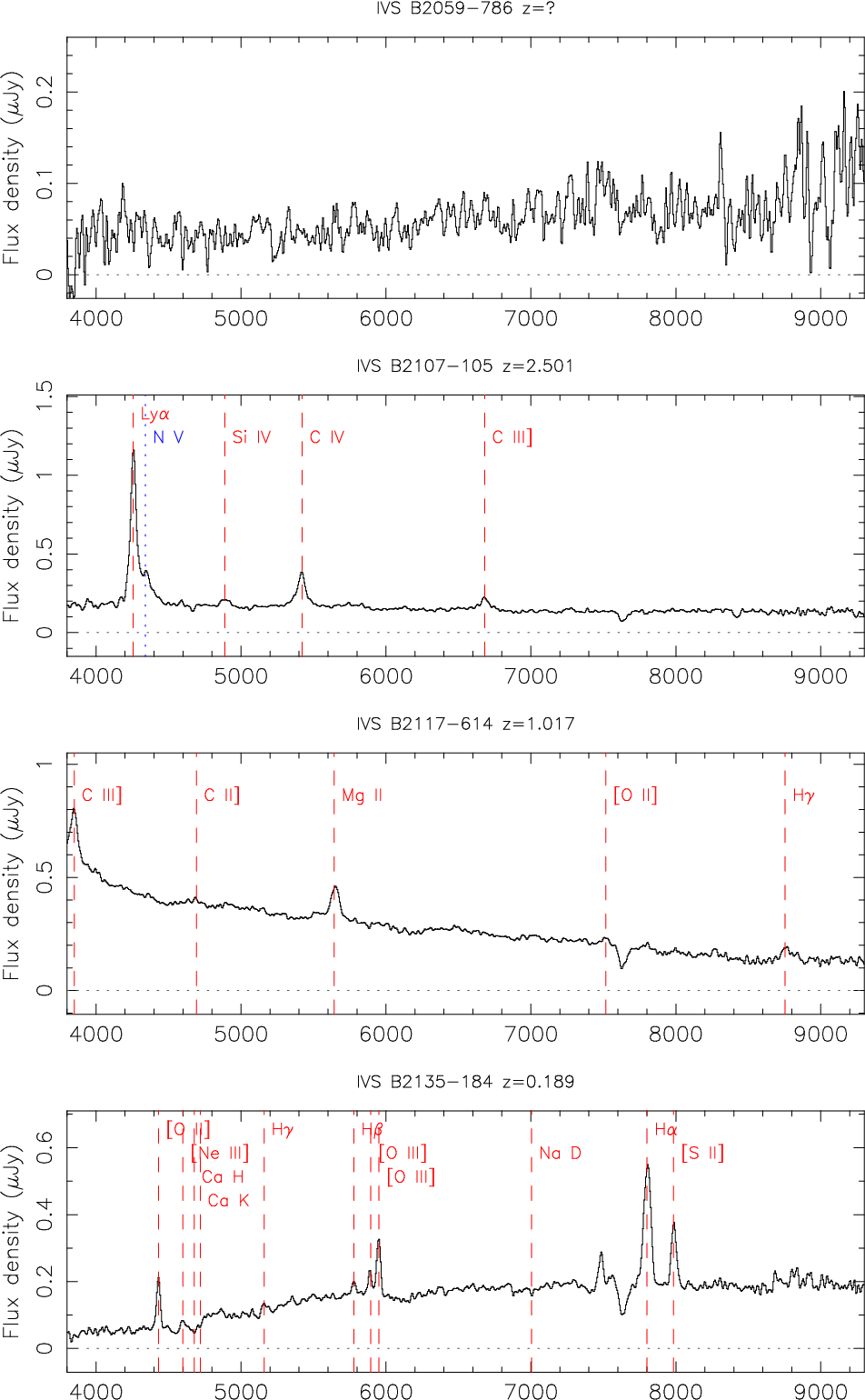}
\caption{}
\end{figure}

\begin{figure}
\figurenum{1 (continued)}
\plottwo{allspec9.ps}{allspec10.ps}
\caption{}
\end{figure}

\begin{figure}
\figurenum{1 (continued)}
\plottwo{allspec11.ps}{blank.ps}
\caption{}
\end{figure}

\begin{figure}
\epsscale{1.0}
\plotone{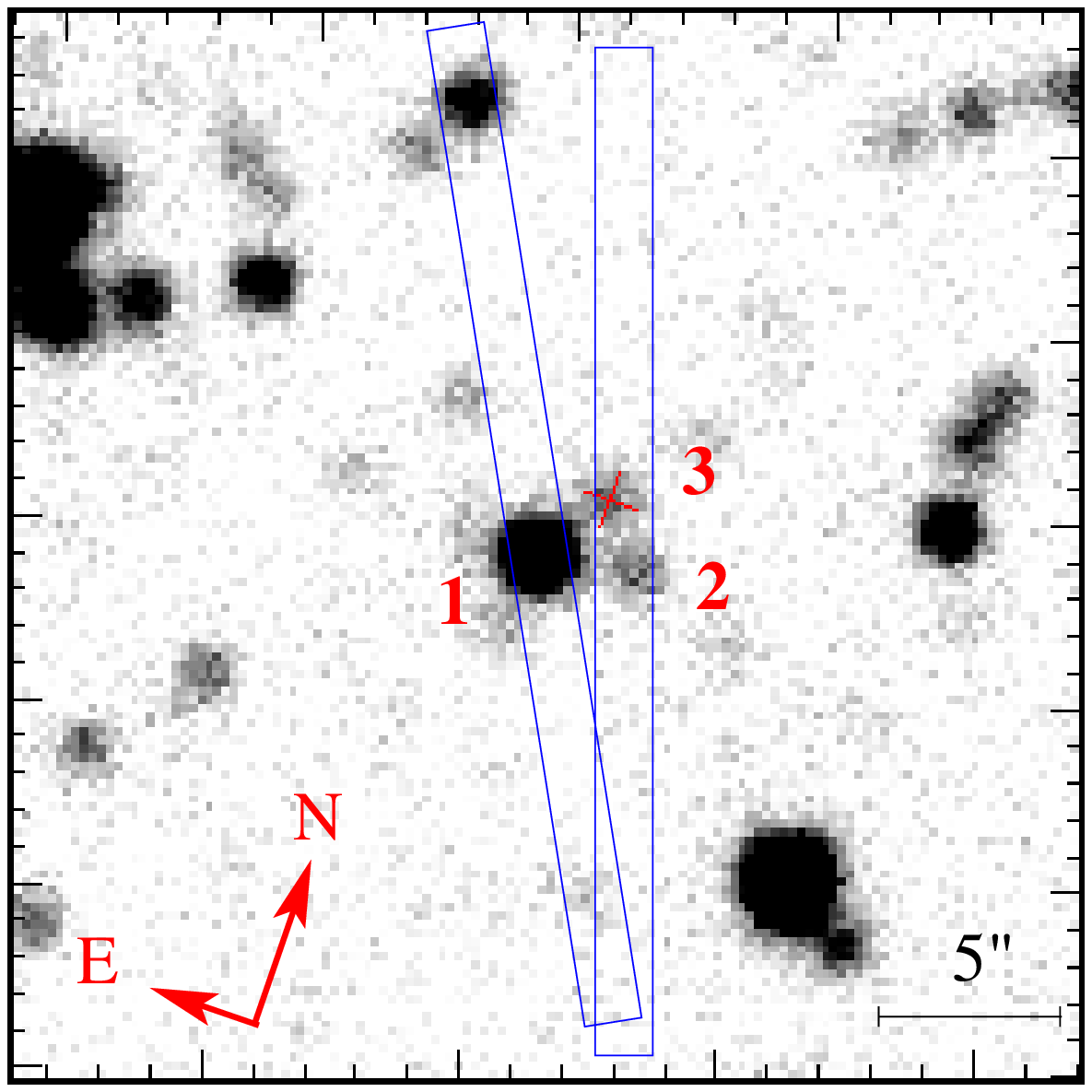}
 \caption{NTT acquisition image of the area near the radio position of the
source IVS B1923+210.  The optical identification is object 3, marked with
a red cross; objects 1 and 2 showed stellar spectra. The two slit
positions used are marked by blue boxes.}
 \end{figure}

\begin{figure}
\epsscale{1.0}
\plotone{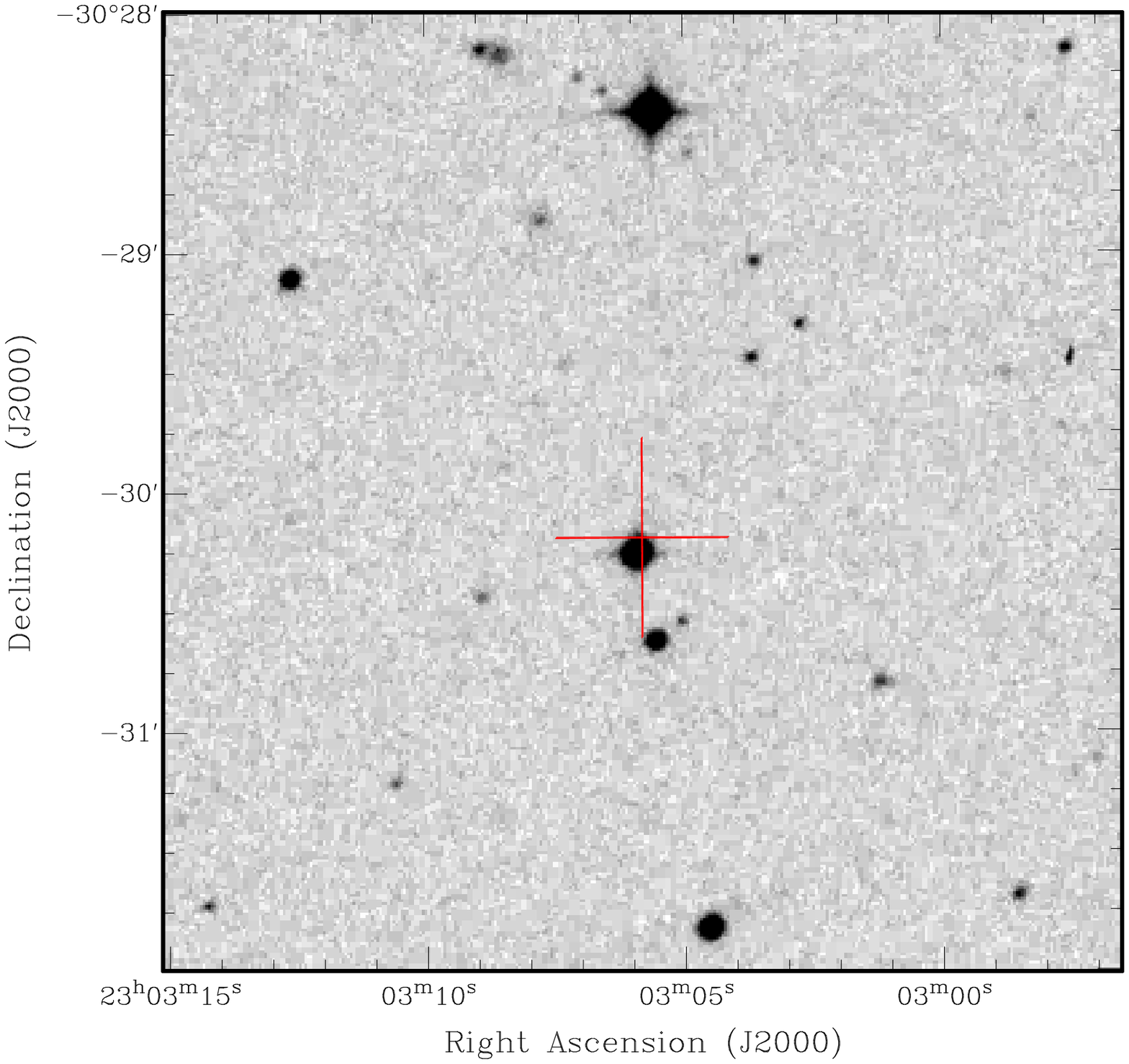}
\caption{Finding chart for the source IVS B2300$-$307 from the  
UK Schmidt B image.  The IVS position is marked with a cross, 
showing how the optical field of the radio source is obscured by the 
foreground star.}
\end{figure}

\end{document}